\documentclass[conference]{IEEEtran}

\usepackage{amsmath,amsfonts}
\usepackage{algorithmic}
\usepackage{graphicx}
\usepackage{textcomp}
\usepackage{xcolor}
\usepackage{pgfplots}
\usepackage{algorithmic}
\usepackage{graphicx}
\usepackage{textcomp}
\usepackage{color,soul}
\usepackage{booktabs}
\usepackage{vcell}
\usepackage{lipsum}
\usepackage{tabularx}
\usepackage{graphicx}
\usepackage{adjustbox}
\usepackage{romannum}
\usepackage{longtable}
\usepackage{array}
\usepackage{caption}
\usepackage{subcaption}
\usepackage{pdflscape}
\usepackage{tabu}
\usepackage{supertabular}
\usepackage{tikz}
\usepackage{pgfplots}
\usepackage{pgf-pie}
\pgfplotsset{compat=newest}
\usepackage{pgfplotstable}
\usepackage{longtable}
\usepackage{multirow}
\usepackage{tabu}
\usepackage{booktabs}

\newcommand\ba[1]{\textcolor{black}{#1}}
\AtBeginDocument{%
  \providecommand\BibTeX{{%
    \normalfont B\kern-0.5em{\scshape i\kern-0.25em b}\kern-0.8em\TeX}}}

\usepackage[ruled, linesnumbered, longend]{algorithm2e}

\begin{document}

\title{Scalable and Programmable Look-Up Table based\\Neural Acceleration (LUT-NA) for\\Extreme Energy Efficiency}

\title{Look-Up Table based Neural Network Hardware}

\author{
\IEEEauthorblockN{Ovishake Sen\textsuperscript{1}, 
Chukwufumnanya Ogbogu\textsuperscript{2}, Peyman Dehghanzadeh\textsuperscript{1},\\Janardhan Rao Doppa\textsuperscript{2}, Swarup Bhunia\textsuperscript{1}, Partha Pratim Pande\textsuperscript{2}, and Baibhab Chatterjee\textsuperscript{1}}
\IEEEauthorblockA{\textit{\textsuperscript{1}
Department of Electrical and Computer Engineering, University of Florida,}
Gainesville, USA.\\
\textit{\textsuperscript{2}School of Electrical Engineering \& Computer Science, Washington State University,}
Pullman, USA.\\
email: ovishake.sen@ufl.edu, c.ogbogu@wsu.edu, p.dehghanzadeh@ufl.edu,\\jana.doppa@wsu.edu, swarup@ece.ufl.edu, pande@wsu.edu, chatterjee.b@ufl.edu
}
}

\maketitle

\begin{abstract}
Traditional digital implementations of neural accelerators are limited by high power and area overheads, while analog and non-CMOS implementations suffer from noise, device mismatch, and reliability issues. This paper introduces a CMOS Look-Up Table (LUT)-based Neural Accelerator (LUT-NA) framework that reduces the power, latency, and area consumption of traditional digital accelerators through pre-computed, faster look-ups while avoiding noise and mismatch of analog circuits. To solve the scalability issues of conventional LUT-based computation, we split the high-precision multiply and accumulate (MAC) operations into lower-precision MACs using a divide-and-conquer-based approach.
We show that LUT-NA achieves up to $29.54\times$ lower area with $3.34\times$ lower energy per inference task than traditional LUT-based techniques and up to $1.23\times$ lower area with $1.80\times$ lower energy per inference task than conventional digital MAC-based techniques (Wallace Tree/Array Multipliers) without retraining and without affecting accuracy, \ba{even on lottery ticket pruned (LTP) models that already reduce the number of required MAC operations by up to 98\%}. Finally, we introduce mixed precision analysis in LUT-NA framework for various LTP models (VGG11, VGG19, Resnet18, Resnet34, GoogleNet) that achieved up to $32.22\times$-$50.95\times$ lower area across models with $3.68\times$-$6.25\times$ lower energy per inference than traditional LUT-based techniques, and up to $1.35\times$-$2.14\times$ lower area requirement with $1.99\times$-$3.38\times$ lower energy per inference across models as compared to conventional digital MAC-based techniques with $\sim$1\% accuracy loss.
\end{abstract}

\section{Introduction}
The surge in data-intensive applications and the need for faster processing have challenged traditional computing architectures, especially during the last decade. Machine learning (ML) applications rely on huge amounts of data for training and inference. Training deep neural networks (DNNs) and handling large datasets entail significant computational power, \ba{chip area}, and memory bandwidth. As the complexity of neural network models increases, the demand for faster processing and efficient memory usage becomes even more critical. To tackle these challenges, recent work has proposed domain-specific accelerators to enable energy-efficient processing of deep neural networks. 
The energy, area, and latency costs become even more important for complex DNN workloads that require millions/billions of MAC operations, \ba{especially with traditional SRAM-based implementations}. This limits their scalability to accelerate relatively large DNN workloads \cite{lu2020benchmark}. This challenge is further amplified when we require energy-efficient computation on resource constrained nodes such as biomedical implants and wearables \cite{chatterjee2023bioelectronic, chatterjee_DnT, Rikky_Seizure, chatterjee2023biphasic}.

To address these challenges \ba{of energy and area efficiency}, in this work, we propose a highly programmable look-up table (LUT) based framework for neural acceleration, to be termed as LUT-NA in the rest of the paper. The LUT-NA framework uses a fast, flexible processing architecture \ba{using pre-computed results, supported by} a novel mapping scheme to overcome the scalability and energy efficiency challenges of existing \ba{LUT-based} architectures. We then show an approximate computing technique with layer-dependent mixed-precision analysis that further reduces the energy and area consumption without losing accuracy \ba{($\sim$1\% degradation)}. All of these are achieved on networks with Lottery-Ticket-Pruning (LTP) that maximally prunes the weights without any significant accuracy loss ($<$1\%) but reduces the number of MAC operations by up to 98\%. Thus, LTP further improves the scalability of LUT-NA to larger DNN workloads. Overall, we achieve a 3.37$\times$ better energy efficiency even in the Pruned Resnet34 model w.r.t. the traditional array multiplier with $<$1\% accuracy loss.
The main contributions of this work are: \begin{itemize}
    \item We demonstrate programmable and scalable LUT-based MAC for neural acceleration (LUT-NA) using \ba{a novel} divide \& conquer approach \ba{that makes LUT-based architectures scalable over several DNN models and bit-resolutions}.
    \item We \ba{analyzed the} 
    data and weight \ba{resolution} requirements for different LTP pruned deep-learning models and, for the first time, analyzed the combined effect of weight pruning (LTP) and LUT-NA on the required number of MAC operations and the \ba{model accuracy}.
    \item We introduce the notions of mixed precision analysis and approximate computing in the LUT-NA framework that further reduce the energy and area consumption with the help of \ba{an optimally chosen mixed-precision} approximate computing architecture with only $\sim$1\% accuracy loss.
    \item We analyzed the hardware efficiency (energy per inference and area consumption) for LUT-NA and approximate/mixed-precision LUT-NA for different deep-learning models.
\end{itemize}

In essence, we develop a scalable fabric for neural acceleration, with \ba{fully programmable weight and data on several DNN models, that achieves up to 29.54$\times$ benefit in area and up to 3.34 $\times$ benefit in energy per inference than traditional LUT based techniques without accuracy loss. With mixed-precision analysis, we achieve $32.22\times$ (VGG11) to $50.95\times$ (Resnet34) lower area with $3.68\times$ (VGG11) to $6.25\times$ (Resnet34) benefit in energy per inference than traditional LUT based techniques, with only $\sim$1\% accuracy loss}.

The remainder of the paper is organized as follows: Section 2 summarizes the related work; Sections 3 and 4 explain the proposed methodology and the bit resolution requirement for activations and weights; Section 5 shows the experimental results; finally, the conclusion and future directions are presented in Section 7.

\section{Related Work}
\ba{The contributions and limitations of state-of-the-art (SoA)} LUT-based machine learning accelerators and mixed precision analysis on the deep learning models are listed below:

\ba{An LUT based framework for} energy-efficient processing in cache support for neural network acceleration was proposed in \cite{ramanathan2020look}. \ba{This work achieves} 1.72$\times$ higher performance and 3.14$\times$ lower energy consumption than SoA processing-in-cache solution with the inception V3 model. However, the authors did not provide the area overhead and details of the performance metrics on their machine-learning models. 
A LUT-based multiplier for a systolic array-based Convolutional Neural Network (CNN) accelerator was proposed in \cite{liu2022lut}. Compared to the reference designs with various conventional multipliers, this work achieves up to 23.34\% and 33.26\% reduction in power consumption and power area product (PAP). However, the authors did not provide much detail about the area overhead and \ba{the bit resolution for weights and activations}. The performance metrics of the CNN model were also not provided.
A programmable LUT-based Processing in Memory (PIM) architecture capable of performing massively parallel Data Encryption using the Advanced Encryption Standard (AES) algorithm was proposed in \cite{sutradhar2021ultra}. Though the authors built their LUT-based technique for encryption purposes,  the method shown achieves 1.8$\times$ higher maximum throughput than an SOA GPU Computing Processor for 17.2$\times$ lower maximum power consumption. However, the area overhead was not provided. Also, the authors did not offer details about the ML models used to evaluate the techniques and performance metrics results.
In \cite{vasquez2021activation}, an activation density-based mixed precision quantization with \ba{pruning was proposed for energy-efficient neural networks}, which achieves approximately 198$\times$ and 44$\times$ energy reductions for VGG19 and ResNet18 architectures, respectively, compared to baseline 16b precision, unpruned models. This approach of activation density-based quantization coupled with \ba{pruning in the individual layers of the model} has approximately 4-5\% test accuracy loss \ba{w.r.t.} the full precision for all layers. Also, the authors did not offer the area overhead for their proposed method, \ba{for the reported} VGG19 and Resnet18 models.

Our proposed method achieves the baseline accuracy of the traditional LUT-based methods and the conventional digital MAC-based techniques without affecting accuracy, \ba{while significantly reducing the energy and area overhead compared to the conventional LUT and digital MAC-based methods}. To evaluate the efficacy of the proposed method across various deep learning models, LTP-pruned VGG11, VGG19, Resnet18, Resnet34, and GoogleNet models were used, and the performance metrics were provided.

\section{Proposed LUT-NA framework}
Our proposed method of LUT-based neural acceleration (LUT-NA) \ba{utilizes a divide and conquer (D\&C) technique to improve the scalability in terms of bit-resolution \cite{dehghanzadeh2023luna}. The hardware cost of traditional LUT-based MAC} increases super-linearly with increasing size of the multipliers. \ba{In fact, if we have two n-bit inputs to the multiplier, the number of traditional LUT entries needs to be $2^{2n}$. However, since one of the inputs (the weight) in a trained DNN is a constant, the number of required LUT entries become $2^{n}$. Using D\&C, the complex multiplications with higher bit-resolution are broken down recursively} into smaller multiplications with low-enough resolution and complexity (in terms of hardware and latency). 
Finally, the solutions of the smaller multiplications are combined to get the desired output of the intended MAC \cite{dehghanzadeh2023luna}.
\ba{However, to ensure acceptable accuracy, the required bit-resolution of the system needs to have a minimum value. In the LUT-NA framework presented in this paper, we also explore the effect of approximate MAC operations while preserving the accuracy of the system ($\sim$1\% degradation w.r.t. the baseline accuracy). For evaluation, we consider LTP-pruned VGG11, VGG19, Resnet18, Resnet34, and GoogleNet models on the CIFAR-10 dataset \cite{10244258}.}

\begin{figure}[htbp]
    \centering
    \includegraphics[width=\linewidth]{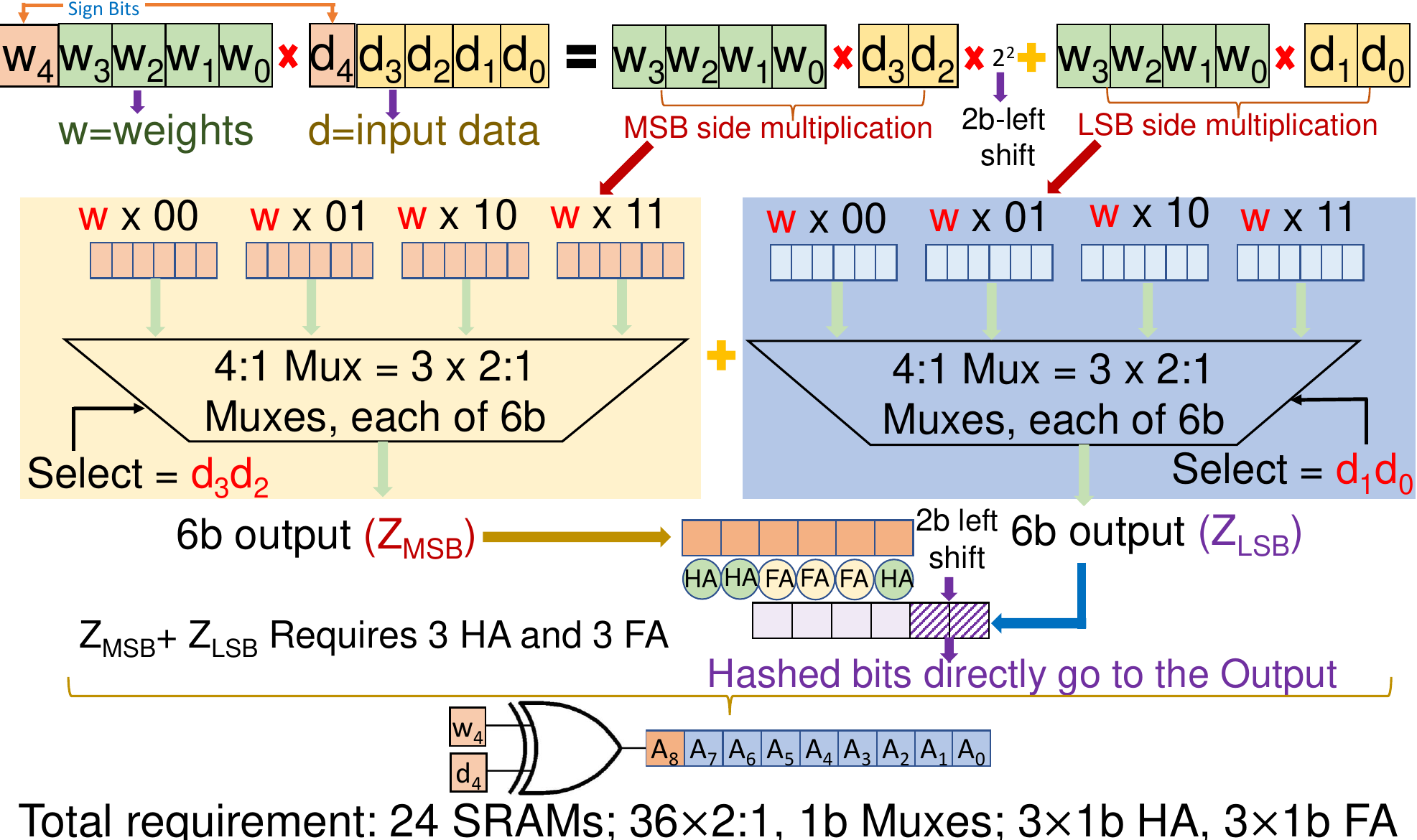}
    \caption{Example 4b $\times$ 4b LUT based multiplier using D\&C}
    \label{fig:lut}
\end{figure}

\subsection{LUT-NA using D\&C approach}
\ba{As an example scenario shown in Fig. \ref{fig:lut}, we explain the LUT based multiplier using D\&C approach for 4b weight ($w$) and 4b data ($d$)}, 
while 1b is reserved extra for the sign bit (we assume a signed magnitude representation).
\ba{We first partition the 4b $\times$ 4b, $w$ $\times$ $d$ multiplication} into two independent 4b $\times$ 2b multiplications. The most significant two bits of $d$ (the MSB side) are used in the first multiplication, while the least significant two bits of $d$ (the LSB side) are used in the second multiplication. The outcome of the MSB side multiplication ($Z_{MSB}$) is shifted by two bits to the left.
The multiplication's final result is then obtained by adding this left-shifted $Z_{MSB}$ to the LSB side multiplication result ($Z_{LSB}$). \ba{Note that the shifting in $Z_{MSB}$ requires no additional hardware, as the adder can inherently be implemented with its input bits shifted to the corresponding locations to represent the shift operation.} The multiplication operation yields the desired result by combining these two partial results. The sign bits are fed into an xor gate and combined into the main result.
For each smaller (4b $\times$ 2b) multiplication, a total of 24b (4 possible 6b results, each requiring 4 $\times$ 6b of storage) is required.
Furthermore, a 4:1, 6b Mux would be employed, equivalent to either three 2:1, 6b Mux instances or eighteen 2:1, 1b Mux instances. Additionally, because of the 2b left shift operation in $Z_{MSB}$ before the addition, the addition of the 6b $Z_{MSB}$ and $Z_{LSB}$ necessitates the use of 3 instances of 1b half adders (HA) and three instances of 1b full adders (FA).

\subsection{LUT-NA using storage-optimized and approximated D\&C approach}
To simplify the hardware, LUT-NA uses a storage-optimized structure on the MSB/LSB side.
\ba{Furthermore, approximate computing concepts are particularly appealing in} neuromorphic applications as neural networks have an inherent error tolerance due to multiple parallel connections from the inputs to outputs \cite{TVLSI19}. 

\begin{figure}[b]
\begin{subfigure}[h]{0.495\linewidth}
\centering
    \includegraphics[width=\linewidth]{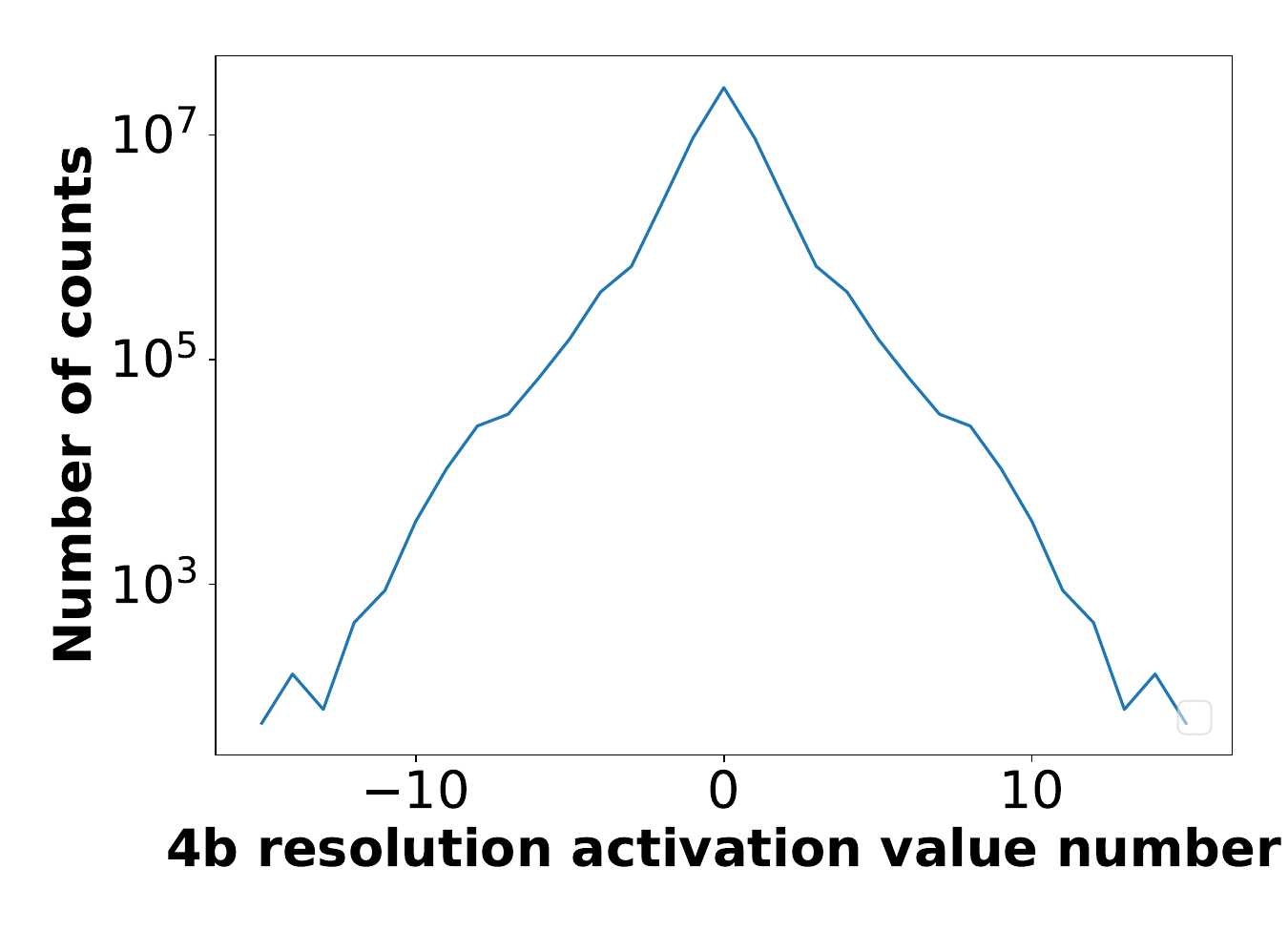}
    \caption{Activation value counts for\\all layers of VGG11 model}
    \label{fig:freq}
\end{subfigure}
\begin{subfigure}[h]{0.495\linewidth}
\centering
    \includegraphics[width=\linewidth]{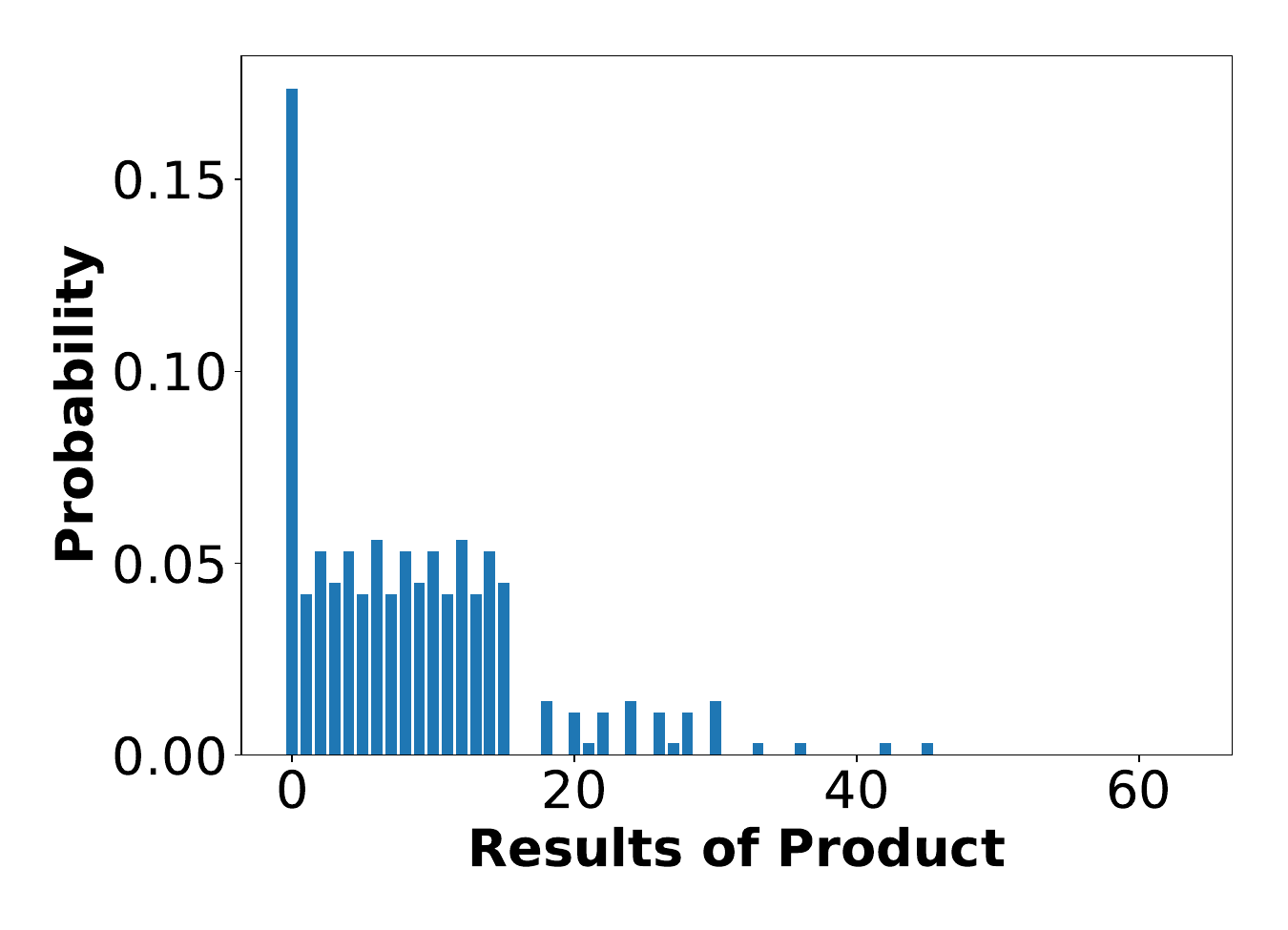}
    \caption{Probability distribution for\\(4b $\times$ 2b) LSB side multiplication
    }
\label{fig:prob}
\end{subfigure}

\caption{The number 0 has a significantly higher probability of occurrence w.r.t. other possible activation values. Hence, the number 0 also has the highest probability of being the 6b product in the LSB side multiplication.}
\label{fig:freq and prob}
\end{figure}

In our approximated D\&C approach in LUT-NA (termed as A-LUT-NA in this paper), if the MSB side is non-zero, the result of the LSB-side multiplication can be approximated to a fixed value, thereby removing the need for any computing hardware for the LSB-side multiplication.
Selecting a fixed $Z_{LSB}$ in this scenario entails determining a specific value that reduces the Hamming distance between the chosen $Z_{LSB}$ and the \ba{original/all possible} $Z_{LSB}$ values. This selection procedure aims to minimize precision loss while maximizing approximation accuracy. The total hardware requirements can be minimized because the LSB side operation only requires a \ba{fixed combination} of 0s and 1s \ba{that does not require any computation/look-up}. To be more precise, the LSB side requires just two bits of storage and does not need any Mux.
For calculating $Z_{MSB}$, $w$ is multiplied with the 2 MSB bits of the input. The combinations of the 2 MSB bits can be 00, 01, 10, and 11. When $w$ is multiplied by 00, the result will be 0, and for that case, only 1 bit (0) is needed for storing the results, which will be connected to all six bits of one of the inputs of the 4:1 Mux. 
When $w$ is multiplied by 01, the result will be $w$. So, storing the four bits of $w$ connected to the 4 LSBs of another distinct 6b input of the 4:1 Mux is sufficient, and the two MSBs will be connected to 0. 
When multiplying $w$ $\times$ 10, just a 0 will be concatenated to the LSB side of the $w$. \ba{Hence, there is no need for any storage,} as the stored 4b result of $w$ $\times$ 01 can be directly connected to \ba{bits b4-b1} of another distinct 6b input of the 4:1 MUX. Meanwhile, this input's MSB and LSB will be connected to 0, \ba{representing the 1b left shift of $w$ $\times$ 01 to obtain $w$ $\times$ 10}. 
Finally, when multiplying $w$ $\times$ 11, it is only necessary to store the 5 MSB of the result. These 5 MSBs will be connected to the corresponding 5  MSBs of the final distinct 6b input of the 4:1 MUX, and the LSB of the 6b input will be connected to the LSB of $w$.
This simplified LUT-NA version requires only 12 SRAM cells,  18 \ba{instances of a} 2:1 1b MUX, 3 HAs of 1b and 3 FAs of 1b.
\ba{Interestingly, the 6b product of (4b $\times$ 2b) LSB-side multiplication cannot produce some specific values between 0-63 and is probabilistically more inclined toward certain other values}. Fig. \ref{fig:freq} illustrates the distribution of activation values (4b resolution) for all the LTP pruned VGG11 model layers for the CIFAR-10 dataset. The number 0 has a significantly higher number of occurrences than other possible activation values. Also, most activation values lie close to the 0 value. Fig. \ref{fig:prob} illustrates the probability distribution for all possible (4b $\times$ 2b) multiplications, where the distribution of the operands follow Fig. \ref{fig:freq}. Fig. \ref{fig:prob} shows that the number 0 has the highest probability of being the 6b product, which motivates us to select $Z_{LSB}$=0.
\ba{Now, setting $Z_{LSB}$ to 0 results in having no extra storage or muxing for $Z_{LSB}$}. Furthermore, nothing needs to be added to $Z_{MSB}$ because $Z_{LSB}$ is estimated to be 0, \ba{which obviates the need for any HA or FA}.\par

\begin{figure}[htbp]
    \centering
    \includegraphics[width=\linewidth]{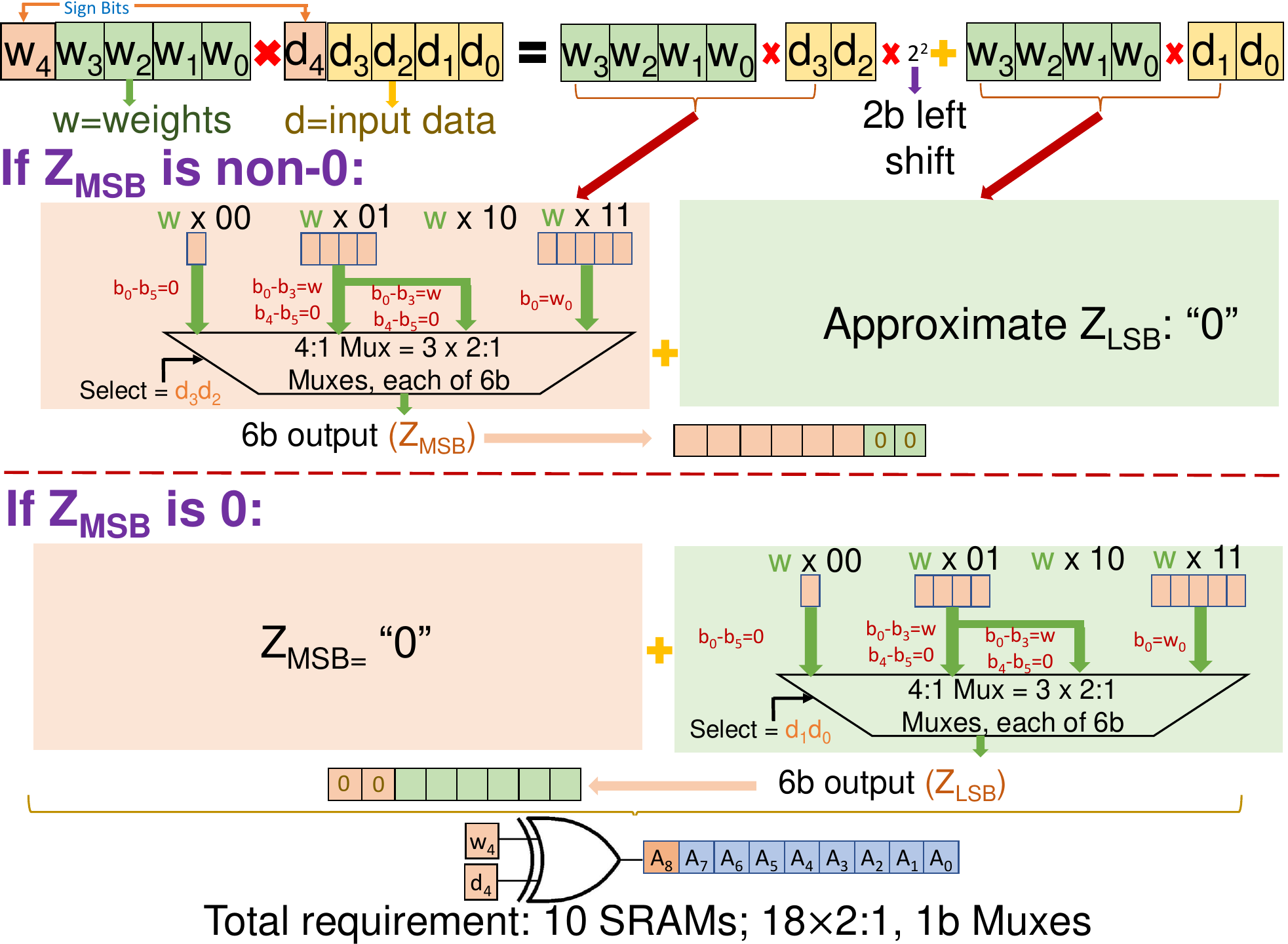}
    \caption{Proposed LUT-NA framework using the storage-optimized and approximated D\&C approach (A-LUT-NA)}
    \label{fig:applut}
\end{figure}

Contrary to all of this, if the MSB side is all zeros, we will consider only the LSB side multiplication. The hardware requirement for this scenario will be the same as that with a non-zero MSB side. As the $Z_{MSB}$ is 0, no extra storage or muxing is required for $Z_{MSB}$, and nothing needs to be added to $Z_{LSB}$. The final storage optimized and approximated D\&C approach requires only 10 SRAM cells and 18 instances of a 2:1 1b MUX. Fig. \ref{fig:applut} illustrates the \ba{4b example for the LUT-NA platform} using the optimized and approximated D\&C approach.

\subsection{Lottery Ticket Pruning}
Pruning is a well-known method aimed at significantly reducing the size of deep neural networks (such as CNNs) without compromising their accuracy \cite{han2015deep} \cite{10244258}. This process typically involves identifying and removing redundant or less important weight parameters from a network. Hence, this leads to less network parameters w.r.t. their unpruned version, which results in a reduction of the required number of MAC operations involved in the neural network. These pruned models can then be deployed on a resource-constrained platform (e.g., LUT-NA) without losing their predictive power. 
In this work, we leverage a state-of-the-art model pruning technique known as Lottery-Ticket-Pruning (LTP) \cite{frankle2018lottery} to improve the scalability of LUT-NA for deep neural networks. The LTP method is an iterative pruning approach that finds highly sparse CNN models with similar or even better test accuracy compared to their unpruned counterparts \cite{liang2021pruning}\cite{han2015deep}. Traditional pruning techniques usually remove weights after training, which may result in information loss, and require more training to recover the accuracy. LTP overcomes this challenge by pruning the model before training, and the pruned model can then be used for inference. Moreover, unlike other pruning methods, models pruned with LTP can be transferred to new datasets, thereby having better generalization \cite{morcos2019one}. 

LTP iteratively trains, prunes, and resets the CNN weights ($\theta$) over $N$ rounds. Each round prunes $p\%$ of the weights that survived the previous round. Algorithm 1 presents the high-level details of the LTP strategy for CNNs used in this work. We start by initializing the CNN weights using Xavier initialization (line 1) \cite{pmlr-v9-glorot10a}, and train the CNN for $E$ epochs. On each iteration, $p\%$ of the CNN weights are pruned based on their magnitude with negligible accuracy drop ($<$1\%) compared to its unpruned version (lines 2-4). Finally, the pruned and trained CNN model (line 6-7) is mapped to the LUT-NA platform for inferencing. Consequently, this yields a significant improvement in energy and area-efficiency, as well as the scalability of LUT-NA to large CNN models by significantly reducing the number of MAC operations. \ba{In this work, we achieve up to 98\% sparsity for the VGG19 model trained using the CIFAR-10 dataset \cite{10244258}, and an average of 68\% sparsity across all 5 CNN models. Consequently, this leads to a corresponding reduction in the total number of MAC operations when executed on LUT-NA, thereby enabling more energy-efficient CNN inference compared to their unpruned counterparts. However, in this paper, we report the additional energy and area benefits that we obtain because of LUT-NA to emphasize the efficacy of the LUT-NA framework in already pruned models.}

\begin{algorithm}
\footnotesize
\caption{Lottery Ticket Pruning (LTP)}
\label{alg:generator}
\SetKwProg{generate}{Function \emph{generate}}{}{end}
\SetKwInOut{KwIn}{Input}
    \SetKwInOut{KwOut}{Output}

    \KwIn{Unpruned CNN model, prune percent ($p\%$)}
    \KwOut{Trained Sparse CNN model}

\textbf{Initialize} $\theta \leftarrow \theta_{initial}$\;
\While{itr $<$ N and accuracy drop $<$ 1\%}{
     \textbf{Train} CNN for E epochs\;
     \textbf{Prune} $p\%$ of $\theta$ based on magnitude\;
     \textbf{Reinitialize} remaining $\theta$ as the new $\theta_{initial}$\;
}
\textbf{Train} sparse CNN for $E$ epochs\;
\textbf{Return} Trained sparse CNN model
\normalsize
\end{algorithm}

\begin{figure}[h]
\begin{subfigure}[h]{0.5\linewidth}
\centering
    \includegraphics[width=\linewidth]{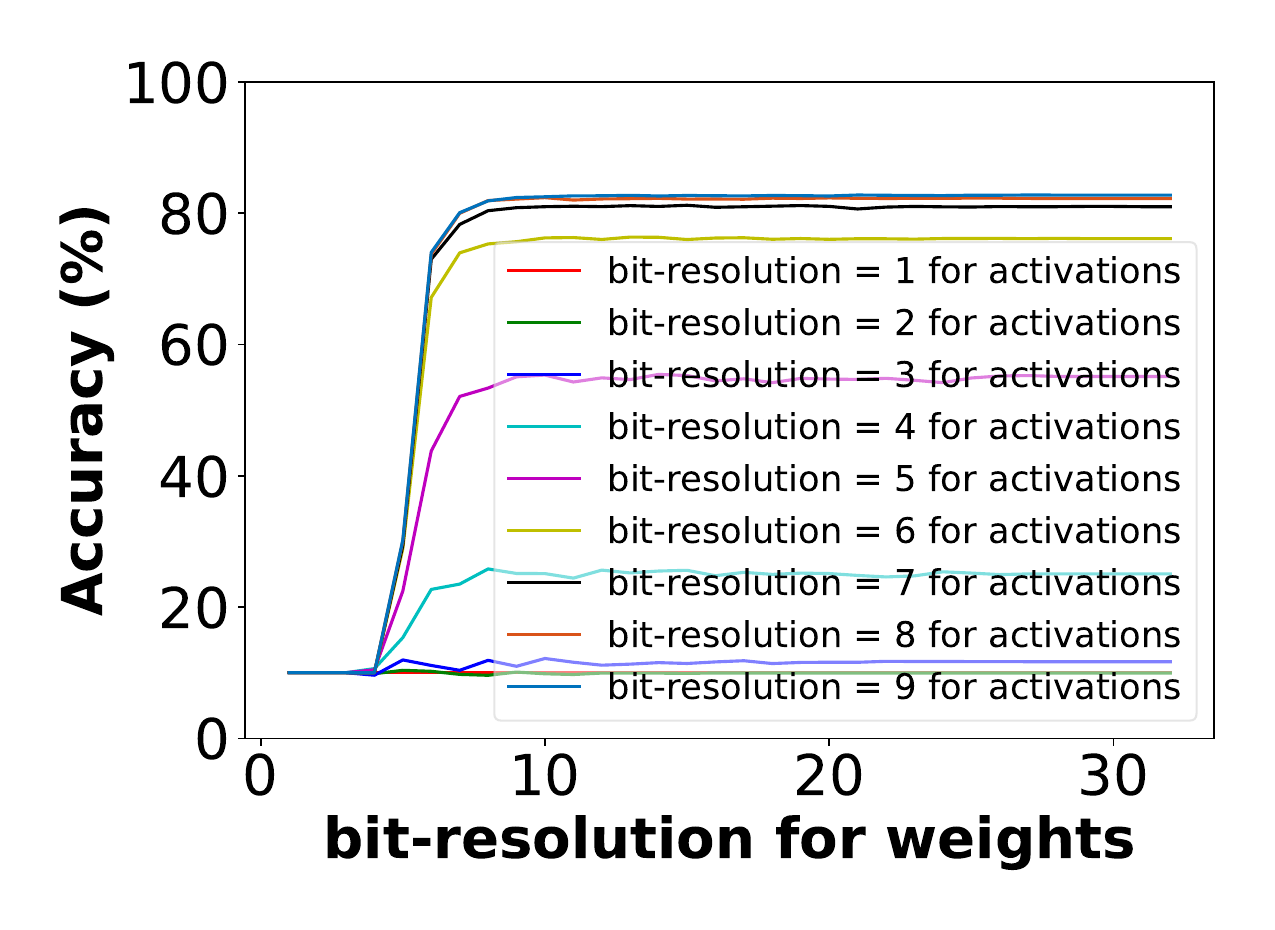}
    \caption{LTP pruned VGG11 model}
\label{fig:vgg111wanda}
\end{subfigure}\hfill
\begin{subfigure}[h]{0.5\linewidth}
\centering
    \includegraphics[width=\linewidth]{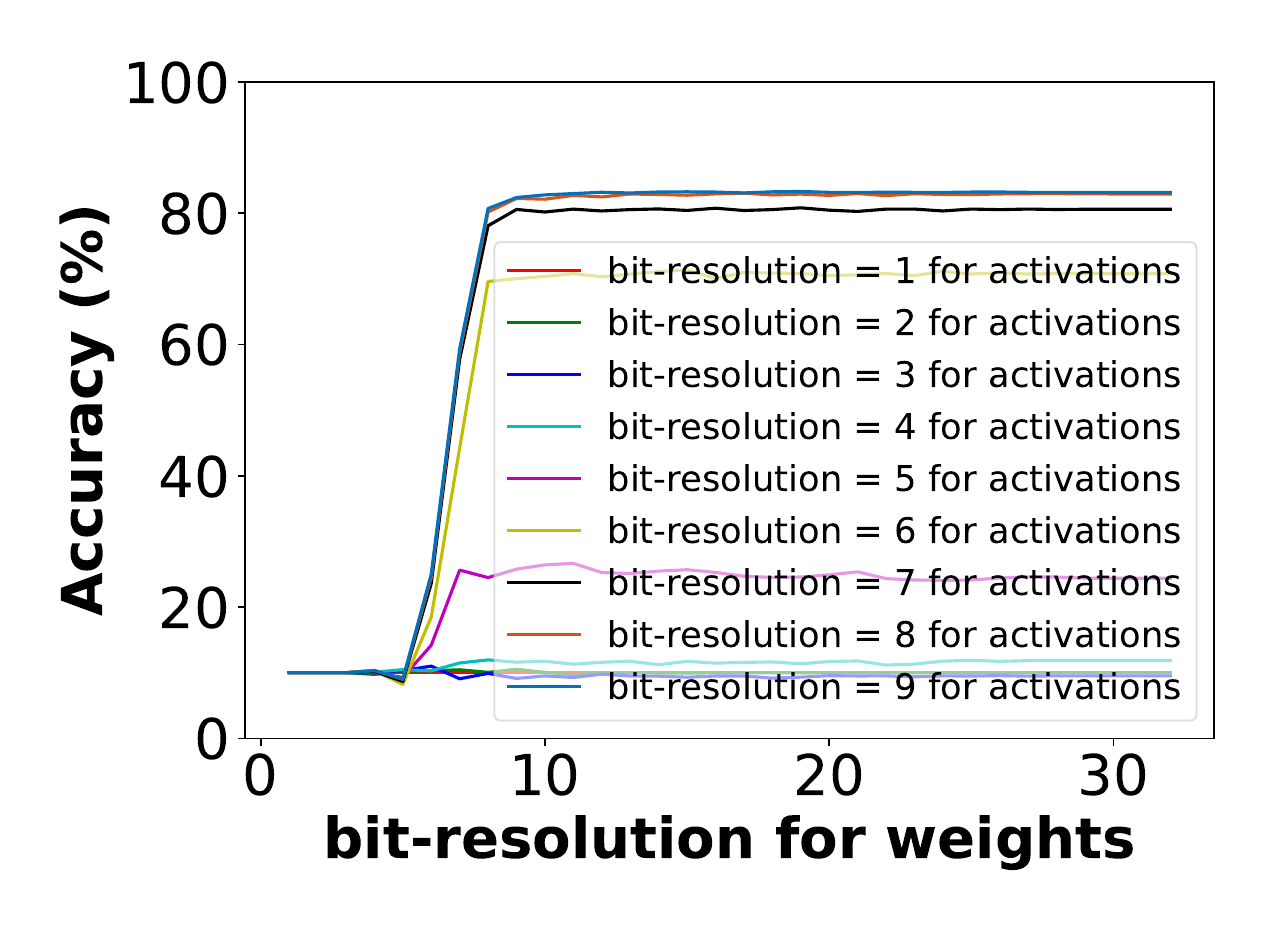}
    \caption{LTP pruned Resnet18 model}
\label{fig:resnet18wanda}
\end{subfigure}\hfill

\caption{Bit-resolution requirement for activations and weight on the LTP pruned VGG11 and Resnet18 models, which illustrates that 8b activations and 9b weights (1 sign bit) are needed to reach the baseline accuracy of the models. }
\label{fig:weightandactivations}
\end{figure}

\begin{figure}[htbp]
\begin{subfigure}[h]{\linewidth}
\centering
    \includegraphics[width=\linewidth]{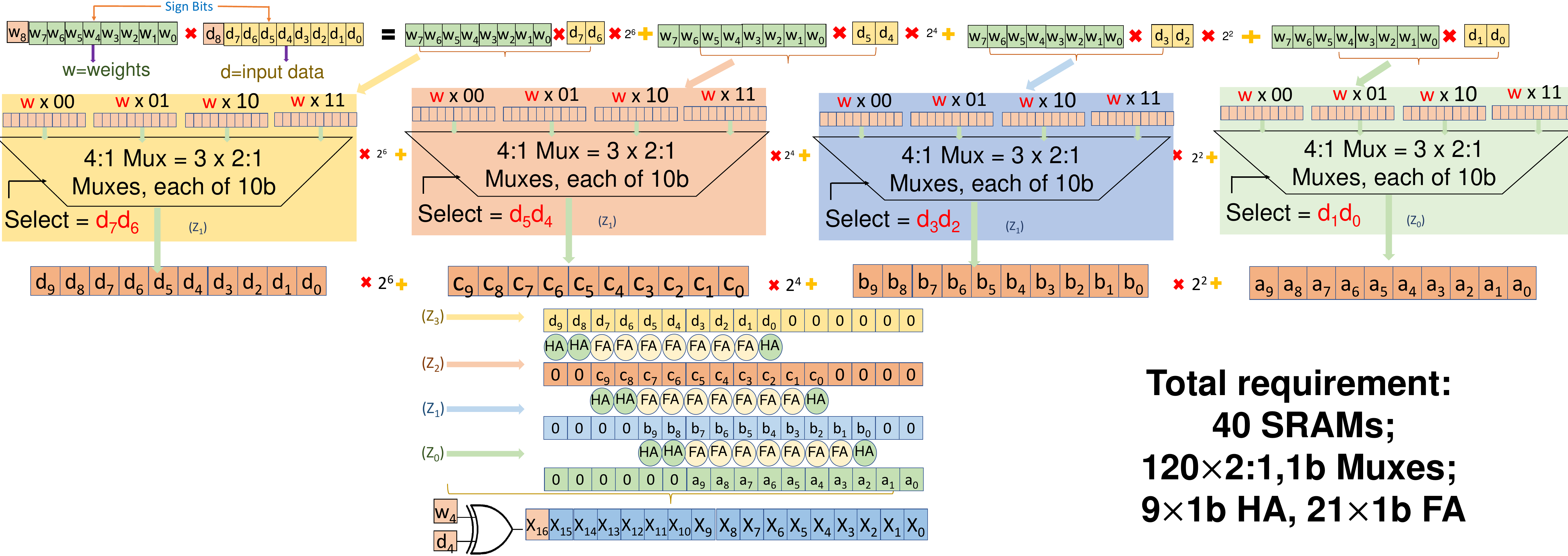}
    \caption{8b LUT-NA using D\&C approach}
\label{fig:9x9l}
\end{subfigure}\hfill
\begin{subfigure}[h]{\linewidth}
\centering
    \includegraphics[width=\linewidth]{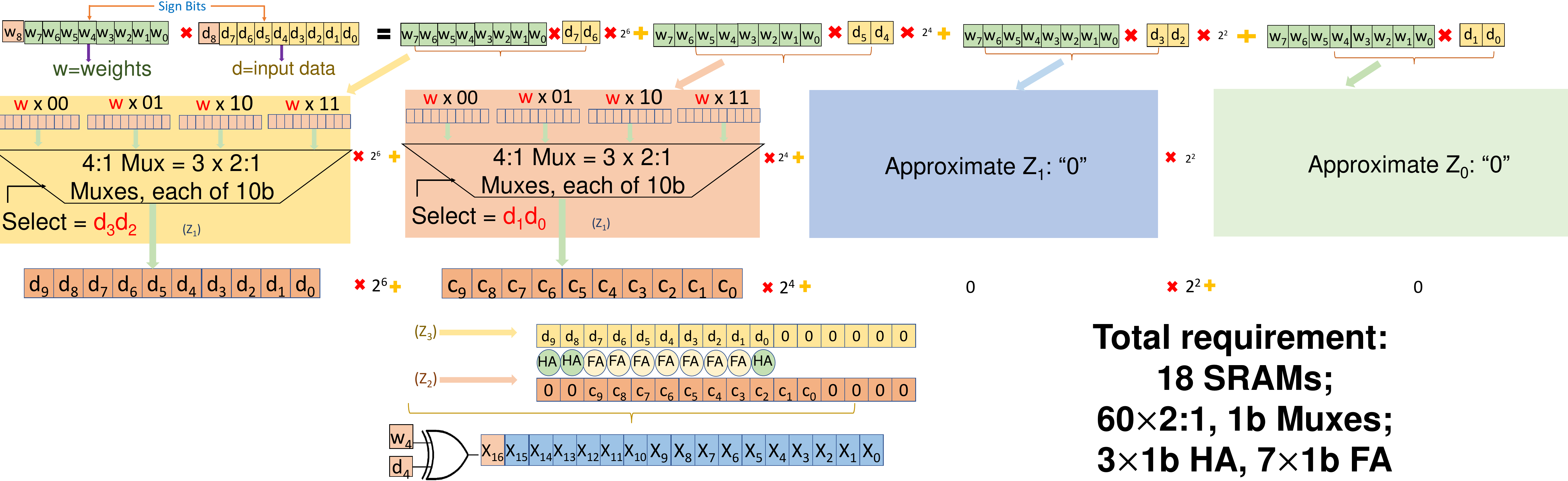}
    \caption{8b LUT-NA using storage-optimized and approximated D\&C}
\label{fig:9x9a}
\end{subfigure}\hfill

\caption{8b LUT-NA used in this paper, utilizing the D\&C approach and storage-optimized and approximated D\&C approach. 1 additional bit is reserved for the sign. }
\label{fig:9x9}
\end{figure}

\section{Bit-resolution requirement for Activations and Weights}
We analyzed the LTP-based pruned models with different activation and weight bits. For this experiment, we used the CIFAR-10 dataset \cite{recht2018cifar}. Fig. \ref{fig:vgg111wanda} and \ref{fig:resnet18wanda} show the VGG11 and Resnet18 pruned model accuracy with \ba{different bit-resolutions of} activations and weights, respectively. Fig. \ref{fig:vgg111wanda} and \ref{fig:resnet18wanda} depict that only 8 bits of activations, including the input and 9 bits of weights, are sufficient to get the baseline accuracy. In this paper, we worked with 9b floating point activations and 9b floating point weights, including the sign bit. 
Our proposed architecture for an 8b D\&C LUT-NA is shown in Fig. \ref{fig:9x9l}, and the LUT-based multiplier using the optimized and approximated D\&C approach is shown in Fig. \ref{fig:9x9a}.

\section{Experimental Results}
\subsection{LUT-NA vs. traditional techniques}
\begin{figure}[tbp]
\begin{subfigure}[h]{0.495\linewidth}
\centering
    \includegraphics[width=\linewidth]{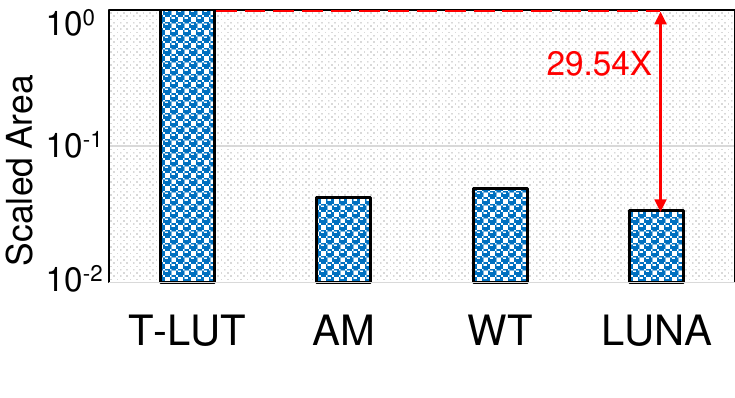}
    \caption{Area consumption analysis}
\label{fig:arealuna}
\end{subfigure}
\begin{subfigure}[h]{0.495\linewidth}
\centering
    \includegraphics[width=\linewidth]{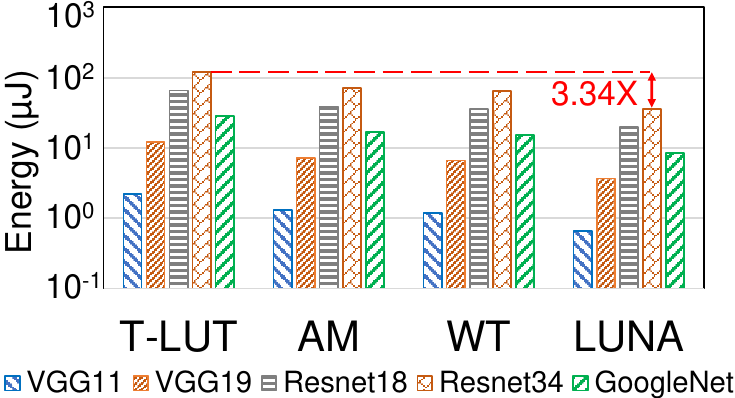}
    \caption{Energy/Inference}
\label{fig:energyluna}
\end{subfigure}

\caption{8b LUT-NA achieves $1.23\times$ lower area and $1.80\times$ lower energy consumption than conventional digital MAC-based methods, as well as $29.54\times$ lower area and $3.34\times$ lower energy per inference than traditional LUT (T-LUT)-based methods.}
\label{fig:areaandenergyluna}
\end{figure}

Using the LTP pruned VGG11, VGG19, Resnet18, Resnet34, and GoogleNet models for the CIFAR-10 dataset, we compared the LUT-NA framework results with conventional methods.
Table \ref{tab:lunavstrad} shows that the LUT-NA framework can be used with only 8b resolution to achieve the accuracy of 32b resolution in traditional digital MAC-based techniques where 1 bit is reserved for the sign bit. Fig. \ref{fig:arealuna} and \ref{fig:energyluna} compare the area and energy consumption (in a standard 45nm CMOS implementation) of 8b LUT-NA with conventional methods. The LUT-NA framework with 8 bit-resolution offers up to $1.23\times$ lower area with $1.80\times$ lower energy consumption than conventional digital MAC-based techniques like Wallace tree (WT) and array multipliers (AM) and a $29.54\times$ lower area with $3.34\times$ lower energy consumption than traditional LUT (T-LUT) based techniques.
Table \ref{tab:lunavstrad}'s 4b LUT-NA fails to reach the baseline accuracy for all models. \ba{However, 4b LUT-NA achieves $>$2-3$\times$ lower power for all models, which is why it is extremely important to investigate approximated D\&C/A-LUT-NA architectures while preserving accuracy as much as possible, which is covered in the next subsection.}

\subsection{Storage-optimized and approximated LUT-NA}
\begin{figure}[tbp]
\begin{subfigure}[h]{0.495\linewidth}
\centering
    \includegraphics[width=\linewidth]{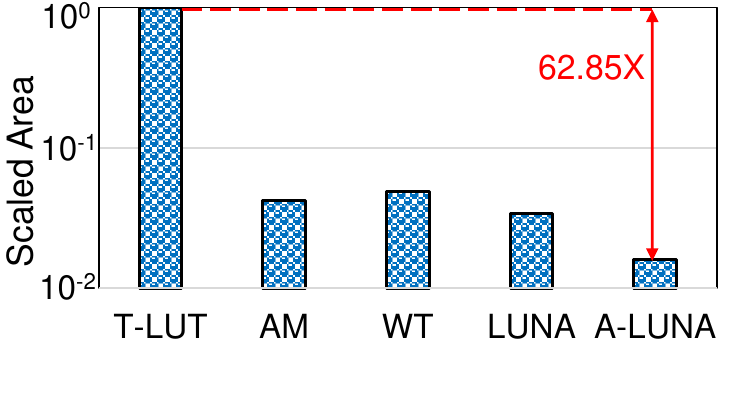}
    \caption{Area consumption analysis}
\label{fig:areaapproluna}
\end{subfigure}
\begin{subfigure}[h]{0.495\linewidth}
\centering
    \includegraphics[width=\linewidth]{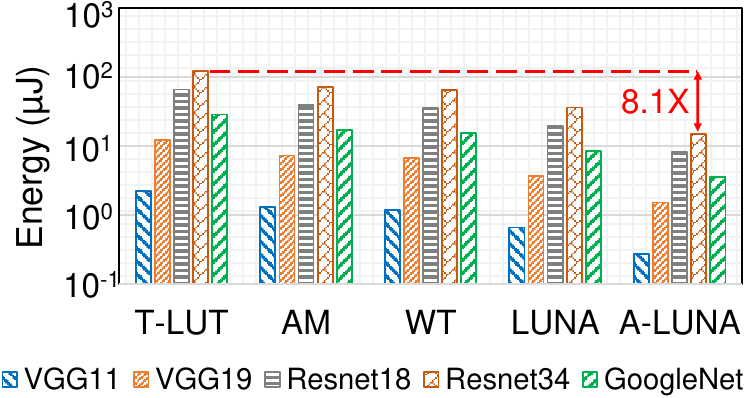}
    \caption{Energy/Inference}
\label{fig:energyapproluna}
\end{subfigure}

\caption{8b Approximate LUT-NA achieves  $2.64\times$ lower area and $4.38\times$ lower energy consumption than conventional digital MAC, with $62.85\times$ lower area and $8.1\times$ lower energy per inference than traditional LUT (T-LUT)-based techniques.  }
\label{fig:areaandenergyapproluna}
\end{figure}

Table \ref{tab:my-table} shows the findings \ba{for storage-optimized and approximated A-LUT-NA framework}, which demonstrates that the approximate 8b LUT-NA implementation results in 5.7\% (GoogleNet) to 26.75\% (Resnet34) loss of accuracy compared to traditional techniques where 1 bit is reserved for the sign bit. However, this approximate 8b LUT-NA implementation \ba{(which has slightly higher amount of hardware as that of 4b LUT-NA)} provides a 15.85\% (VGG11) to 38.55\% (VGG19) improvement in accuracy over the 4b LUT-NA implementation, \ba{while the energy consumption in the 8b A-LUT-NA implementation is $>$2$\times$ better than the original 8b LUT-NA}.
Fig. \ref{fig:areaapproluna} and \ref{fig:energyapproluna} compare the area and energy consumption of A-LUT-NA to LUT-NA and conventional methods. The approximate 8b LUT-NA framework offers up to $2.64\times$ lower area with $4.38\times$ lower energy consumption than conventional digital MAC-based techniques like WT and AM, and up to $62.85\times$ lower area with $8.1\times$ lower energy consumption than T-LUT techniques.
However, A-LUT-NA still has 5.7\% (GoogleNet) to 26.75\% (Resnet34) accuracy loss \ba{as compared to the baseline, as shown in Table \ref{tab:my-table}. Interestingly, the accuracy loss} occurs in various different layers of the network, as a function of the network. That's why we demonstrate a mixed-precision based approach in the next section - where certain layers in the network use the full 8b resolution, and certain other layers use approximated D\&C based LUT-NA to reduce the hardware cost.

\begin{table*}[htbp]
\small
\centering
\caption{Resolution, Accuracy, and Energy per Inference in LUT-NA framework: 8b LUT-NA achieves the baseline accuracy of 32b traditional digital MAC-based techniques on the LTP pruned deep learning models where 1 bit is reserved for the sign.}

\label{tab:lunavstrad}
\begin{tabular}{@{}cccccc@{}}
\toprule
\textbf{Model Name} & \textbf{\begin{tabular}[c]{@{}c@{}}Baseline: 32b traditional\\digital MAC accuracy (\%)  \end{tabular}} & \textbf{\begin{tabular}[c]{@{}c@{}}8b LUT-NA\\ accuracy (\%)\end{tabular}} & \textbf{\begin{tabular}[c]{@{}c@{}}8b LUT-NA\\ energy/inference ($\mu$J)\end{tabular}} & \textbf{\begin{tabular}[c]{@{}c@{}}4b LUT-NA\\ accuracy (\%)\end{tabular}} & \textbf{\begin{tabular}[c]{@{}c@{}}4b LUT-NA\\ energy/inference ($\mu$J)\end{tabular}} \\ \midrule
VGG11 & 82.79 & 82.73 & 0.658 & 55.93 & 0.197 \\
VGG19 & 87.17 & 86.73 & 3.67 & 37.93 & 1.10 \\
Resnet18 & 83.28 & 83.13 & 19.6 & 24.39 & 5.87 \\
Resnet34 & 84.66 & 84.98 & 36.1 & 36.13 & 10.8 \\
GoogleNet & 80.75 & 80.58 & 8.51 & 58.99 & 2.54 \\ \bottomrule
\end{tabular}
\normalsize
\end{table*}

\begin{table*}[htbp]
\small
\centering
\caption{Accuracy and Energy per Inference in Approximate 8b LUT-NA (A-LUT-NA) and Mixed-Precision LUT-NA: A-LUT-NA achieves 5.7\%-26.75\% lower accuracy than a 32b traditional digital MAC, with $\sim$2.4$\times$ lower power than 8b LUT-NA. Mixed-Precision (8b LUT-NA for certain layers, A-LUT-NA for certain other layers) achieves $\sim$1\% accuracy loss than baseline.}

\label{tab:my-table}
\begin{tabular}{@{}ccccccc@{}}
\toprule
\textbf{\begin{tabular}[c]{@{}c@{}}Model\\ Name\end{tabular}} & \textbf{\begin{tabular}[c]{@{}c@{}}8b A-LUT-NA\\accuracy   (\%)\end{tabular}} & \textbf{\begin{tabular}[c]{@{}c@{}}8b A-LUT-NA\\energy/Inference ($\mu$J)\end{tabular}}  & \textbf{\begin{tabular}[c]{@{}c@{}}8b A-LUT-NA\\accuracy loss\\w.r.t. baseline (\%)\end{tabular}} & \textbf{\begin{tabular}[c]{@{}c@{}}Mixed-Precision\\LUT-NA\\accuracy (\%)\end{tabular}} & \textbf{\begin{tabular}[c]{@{}c@{}}Mixed-Precision\\LUT-NA energy/\\Inference ($\mu$J)\end{tabular}} & \textbf{\begin{tabular}[c]{@{}c@{}}Mixed-Precision\\ LUT-NA accuracy loss\\w.r.t. baseline (\%)\end{tabular}} \\ \midrule
VGG11 & 71.78 & 0.271 & 11.01 & 82.27 & 0.6 & 0.52\\
VGG19 & 76.48 & 1.51 & 10.69 & 85.98 & 2.61 & 1.19\\
Resnet18 & 56.53 & 8.10 & 26.75 & 82.61 & 11.2 & 0.67\\
Resnet34 & 63.4 & 14.9 & 21.26 & 83.55 & 19.3 & 1.11\\
GoogleNet & 75.05 & 3.51 & 5.7 & 79.21 & 5.18 & 1.54\\
\bottomrule
\end{tabular}
\normalsize
\end{table*}

\begin{figure}[htbp]
\begin{subfigure}[h]{0.48\linewidth}
\centering
    \includegraphics[width=\linewidth]{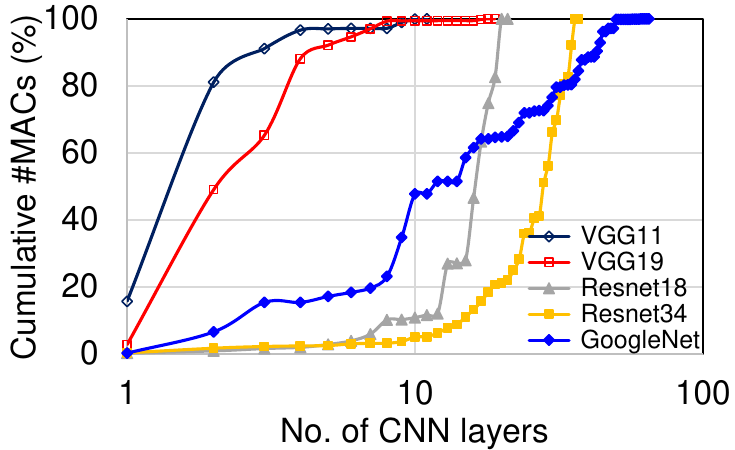}
    \caption{Cumulative no. of MACs\\(\%) for CNN layers}
\label{fig:cumulativemac}
\end{subfigure}
\begin{subfigure}[h]{0.48\linewidth}
\centering
    \includegraphics[width=\linewidth]{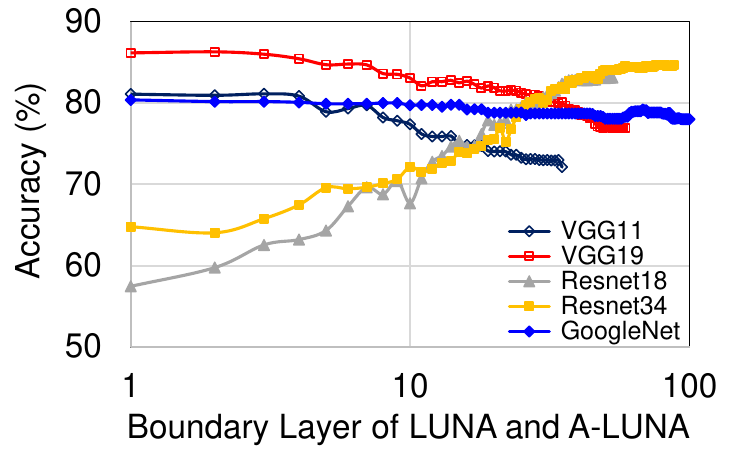}
    \caption{Selecting the boundary layer between LUT-NA and A-LUT-NA 
    }
    \label{fig:mixedfigure}
\end{subfigure}

    \caption{Depending on the layer-wise Cumulative no. of MACs for different CNNs, the boundary between full precision (8b) and approximate operations are selected.
    }
\label{fig:cumulative and mixed figure}
\end{figure}

\begin{figure}[htbp]
\begin{subfigure}[h]{0.495\linewidth}
\centering
    \includegraphics[width=\linewidth]{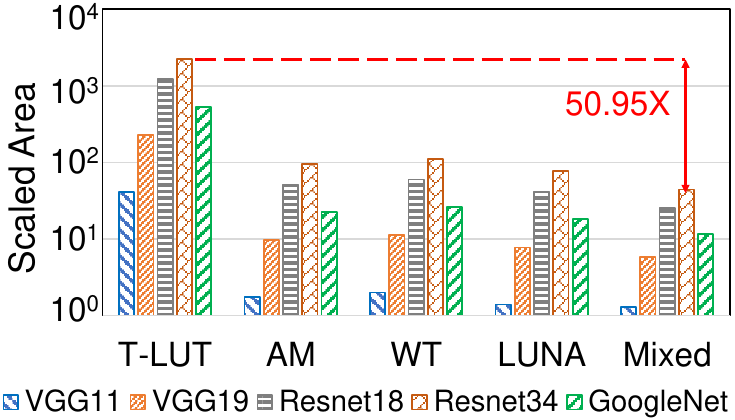}
    \caption{Area consumption analysis}
\label{fig:areamixedluna}
\end{subfigure}
\begin{subfigure}[h]{0.495\linewidth}
\centering
    \includegraphics[width=\linewidth]{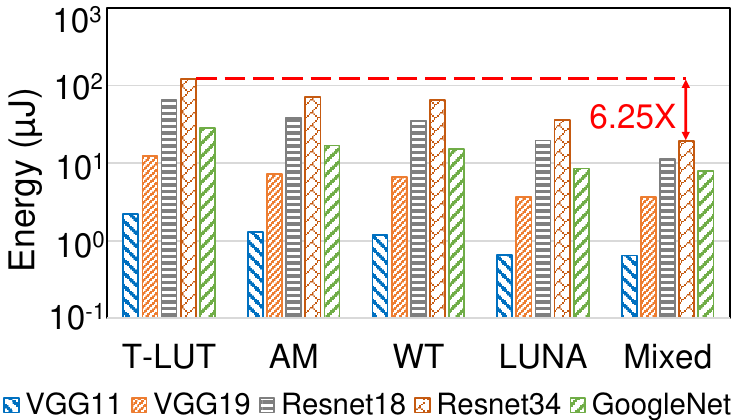}
    \caption{Energy/Inference}
\label{fig:energymixedluna}
\end{subfigure}

\caption{Mixed-Precision LUT-NA achieves  $1.35\times$-$2.14\times$lower area and $1.99\times$-$3.38\times$ lower energy consumption than conventional digital MAC, with $32.22\times$-$50.95\times$ lower area and $3.68\times$-$6.25\times$ lower energy consumption than traditional LUT (T-LUT)-based techniques. }
\label{fig:areaandenergymixedluna}
\end{figure}

\subsection{Mixed precision with LUT-NA and A-LUT-NA}
Fig. \ref{fig:cumulativemac} illustrates the cumulative percentage of MACs as a function of the CNN layers in the LTP pruned VGG11, VGG19, Resnet18, Resnet34 and GoogleNet models. The LTP pruned VGG11, VGG19, and GoogleNet models have more \%-of MACs in the initial layers compared to the Resnet18 and Resnet34 models. Hence, for Resnet18 and Resnet34, it makes sense to use full precision LUT-NA in the initial `n' layers, and A-LUT-NA in the later layers to save more hardware. As we plot the accuracy of these models as a function of the boundary layer between LUT-NA and A-LUT-NA in Fig. \ref{fig:mixedfigure}, the accuracy keeps increasing as a function of `n'.  On the other hand, for VGG11, VGG19, and GoogleNet models, it is more useful to employ A-LUT-NA in the initial `n' layers and full-precision LUT-NA in the later layers to save hardware. The accuracy keeps decreasing for these models as a function of `n' in Fig. \ref{fig:mixedfigure}. In fact, if we employ 8b LUT-NA for the initial 36 and 48 layers for the Resnet18 and Resnet34 models and 8b A-LUT-NA for the subsequent layers, and if we employ 8b A-LUT-NA to the initial 1, 4, and 69 layers for the VGG11, VGG19, and GoogleNet models, respectively, and 8b LUT-NA to the subsequent layers, we can achieve $\sim$1\%  accuracy loss w.r.t. 32b traditional digital MAC as shown in Table \ref{tab:my-table}.
Fig. \ref{fig:areamixedluna} and \ref{fig:energymixedluna} compare the mixed precision implementation's area and energy consumption to LUT-NA and conventional methods. The mixed precision implementation achieves $1.35\times$ (VGG11) to $2.14\times$ (Resnet34) lower area requirements than conventional digital MAC-based techniques and $32.22\times$ (VGG11) to $50.95\times$ (Resnet34) lower area requirements than traditional LUT-based techniques. Also, this mixed precision implementation obtains $1.99\times$ (VGG11) to $3.38\times$ (Resnet34) lower energy consumption than conventional digital MAC-based techniques and $3.68\times$  (VGG11) to $6.25\times$ (Resnet34) lower energy consumption than T-LUT techniques. Note that for LUT-NA or A-LUT-NA implementations, all models scale in the same manner (that is why for Fig. \ref{fig:areaandenergyluna} and \ref{fig:areaandenergyapproluna}, the area plots are not shown separately for each model). On the other hand, for mixed-precision analysis, since each model will have a different percentage of layers implemented using LUT-NA and A-LUT-NA, each model will benefit by a different amount.

\ba{Resnet18, and Resnet34 models benefit the most in terms of accuracy from the mixed-precision analysis, since these are larger models that had significant accuracy degradations in the A-LUT-NA implementation. Interestingly, Resnet18 and Resnet34 also exhibit 43\% and 47\% lower energy per inference with mixed-precision as compared to full 8b precision, as these models benefited from implementing A-LUT-NA on a higher concentration of MACs toward the later part of the network (please refer to Fig. \ref{fig:cumulativemac}).}

\section{Conclusions \& Future Work}
We present a \ba{fully programmable LUT-based Neural Accelerator (LUT-NA) framework that employs a divide and conquer technique by splitting high-precision MAC operations into lower-precision MACs, to improve the scalability of traditional LUT-based methods. LUT-NA demonstrates $29.54\times$ lower area and $3.34\times$ lower energy per inference than traditional LUT-based techniques, and $1.23\times$ lower area with $1.80\times$ lower energy per inference w.r.t. conventional digital MAC due to simple, scalable look-up based hardware. Utilizing the layer-wise information of the number of MAC operations in a network, we introduced mixed precision analysis in the LUT-NA framework that employs approximate computing in the highest possible number of MACs with only $\sim$1\% accuracy loss. This method achieves $32.22\times$-$50.95\times$ lower area across various CNN models (VGG11, VGG19, Resnet18, Resnet34 and GoogleNet), with $3.68\times$-$6.25\times$ lower energy per inference than traditional LUTs, and $1.35\times$-$2.14\times$ lower area with $1.99\times$-$3.38\times$ lower energy per inference across models w.r.t. digital MACs.}
In future, along with image classifications, we will investigate the proposed LUT-NA framework on language models, as well as multivariate, numerical, and categorical datasets. We will also explore the effects of bit-level operations (such as masking) on the individual neuron output in the LUT-NA framework, and shall explore extending the LUT-NA concepts to SRAM as well as non-volatile memory based processing in-memory.

\bibliographystyle{IEEEtran}
\bibliography{1_sample-authordraft}

\end{document}